\documentclass[
 reprint,
 amsmath,amssymb,
 aps,
 showkeys
]{revtex4-1}

\usepackage{graphicx}
\usepackage{dcolumn}
\usepackage{bm}
\usepackage{hyperref}

\begin{document}

\title{Modular architecture of large-scale optical interferometers for sparsely populated input}

\author{Mikhail Yu.~Saygin}
 \email{saygin@physics.msu.ru}
 
\affiliation{
 Quantum Technologies Center and Faculty of Physics, M.V.Lomonosov Moscow State University, GSP-1, Leninskie gory, Moscow 119991  Russian Federation
}


\begin{abstract}
Today,  the realization of large optical interferometer schemes  is necessary for many sophisticated information processing algorithms. 
In this work, we propose a modular interferometer architecture possible  when the number of input channels of a transformation populated with signals is much lesser than the total number of its channels.
The underlying idea is to split the  transformation into two stages. 
In the first stage, the  signals undergo scattering 
among multiple channels of separate optical schemes,
while in the second stage, they are made to interfere with each other, thus, we call it the mixing stage.
This way, large-scale transformations that technically challenging to fabricate even by means of the mature integrated photonics technology, can be realized using multiple low-depth optical circuits, yet at the expense of single interconnection between the scattering and mixing parts of the interferometer.
Although non-universal, 
as witnessed by 
the parameters set to define the interferometer, the architecture provides a  viable approach to construction of large-scale interferometers.

\end{abstract}

\pacs{Valid PACS appear here}
\maketitle

\section{\label{sec:intro}Introduction}

Nowadays, conducting research in optics and creation of 
new optical devices very often require sophisticated optical schemes implementing transformations between multiple channels.
Multiport interferometers represents a general class of linear optical devices that are necessary, 
for example, in  information processing and transfer systems~\cite{Soref} and  
synthesis of spatial optical fields~\cite{Miller}.
In recent years, the interest to multiport optical interferometers
has been stimulated by research of the quantum optics community,
owing to their importance in implementations of 
algorithms~\cite{Aaronson,Politi,OBrienScience}.

To implement  interferometer transformations with standard optical elements,
methods have been devised that enable decomposition of any linear transform into arrays  of elementary blocks of beam-splitters and phase-shifters~\cite{Reck,Clements}.
In particular,  arbitrary $N$-by-$N$ unitaries 
can be decomposed into a planar mesh of $N(N-1)/2$ blocks with 
the minimal depth of the mesh is $N$~\cite{Clements}.
Because of the stringent requirements of high stability and precise control,
universal large-scale optical schemes are hard to implement with conventional bulk optics~\cite{Politi,OBrienScience}.

Currently,  integrated photonics is considered a perspective way to fabricate sophisticated optical schemes,
enabling many functional  elements to be allocated on a compact footprint of a stable photonic chip~\cite{Tanizawa,DeepLearning}, with the possibility to 
co-integrate electronic circuitry into it~\cite{OptoElectronics}.
The standard element of integrated photonic circuits is the two-port Mach-Zehnder interferometer (MZI),
that can serve as a variable  beam-splitter, provided that 
the phases of arms in between the balanced beam-splitters 
constituent the MZI can be manipulated~\cite{Englund}.
With this elements,
integrated planar chips, 
containing from some tens~\cite{Tanizawa} to hundreds of MZIs, have been demonstrated~\cite{DeepLearning}.

Despite the maturity of integrated photonics,  technical obstacles exist that make creation of large optical circuits difficult.
One of the reason is the errors that more often occur at fabrication of large meshes of  elements.
Moreover, the necessity to embed  thermo- and electro-optical elements into the chip to achieve its reconfigurability elevate complexity 
requires extra area budget for wiring and packaging. 
In addition, tuning large optical chips is generally much harder and may be time-consuming, due to  optimization algorithms greedy on computational resources.
In this regard, novel architectures to building  large linear schemes  
capable to overcome the existing technical hurdles is of paramount interest nowadays.

In order to devise more efficient interferometer architectures, 
one can use specifics of the tasks of interest.
In this article, we leverage the fact that often
only a fraction inputs in the interferometer transformation is reserved for signals,
as, for example, in the trivial case of  $M$-to-$N$ splitters and switches with $M\ll{}N$~\cite{Soref}.
A more prominent example of this comes from quantum information 
with the notion of bosom sampling~\cite{Aaronson,Laing,Lund}.
Because this algorithm is strongly believed to be intractable on a classical computer~\cite{Aaronson,Laing,Scheel}, 
the algorithm becomes one of the major contender in the quest to demonstrate  computational advantage of the quantum devices over their classical counterparts.
The optical realizations of this algorithm requires multichannel unitaries,
used to sample photon probability distribution at the output
from a bunch of single photons at the input.
Importantly, the number of input ports occupied by signal photons should scale as $\sim\sqrt{N}$, with $N$ being the number of channels in the unitary,
since today strict prove of computational complexity exists only for this case.
However, to beat modern classical computers one needs at hand  
unitaries of $N\sim1000{}$~\cite{Laing},
which is extremely challenging to  implement currently on a single integrated photonic circuit.

In this work, we suggest an approach to implementing large-scale  
interferometers, that exploits sparsity in the number of input ports occupied by signals. 
To arrive at this architecture, we used the idea of splitting the multichannel transformation into
the scattering stage, in which the signal fields spread among multiple channels separetely without interfering with the others, followed by the mixing stage, that makes the fields to interfere with each other.
Although non-universal, in that it is not capable to yield arbitrary unitary transform, the  interferometer  architecture represents two layers of 
transformation blocks with only inter-layer connection. 
The method is more technically  viable,
since each of the transformation blocks can be fabricated and tested separately,
and the depth of the overall  device turns out to be much smaller, than that of the generic architecture.

\section{\label{sec:splitting}Splitting transformation into scattering and mixing stages}

An $N$-port linear optical device is characterized by an $N$-by-$N$ matrix $U$
that relates the field amplitude at the input, $a_j$,  with the amplitudes at the output, $b_i$, via:
    \begin{equation}\label{eqn:linearTransform}
        b_i=\sum_{j=1}^{N}U_{ij}a_j,\quad(i=\overline{1,N}),
    \end{equation}
Assuming the  transformation is lossless, 
the matrix $U$ is unitary that yields the relation ot its elements:
$\sum_{j=1}^NU_{mj}U_{nj}^{\ast}=\delta_{mn}$.
It is known that  general unitary matrices of dimension $N$ are parametrised by $N^2$ real parameters~\cite{Jarlskog}.
However,  optical schemes that are used to represent linear unitary transforms 
some phase parameters can be omitted, 
since usually a multiport transform is defined up to global phase.
For example, the decomposition methods  by Reck~\cite{Reck} and Clements~\cite{Clements}, having triangular and rectangular arrangement, respectively,
enable to express linear unitaries in the form of planar arrays of $N(N-1)/2$  beam-splitters, each of which is characterized by $2$ real parameters, giving 
$N^2-N$ real parameters

\begin{figure}
    \centering
    \includegraphics[width=0.3\textwidth]{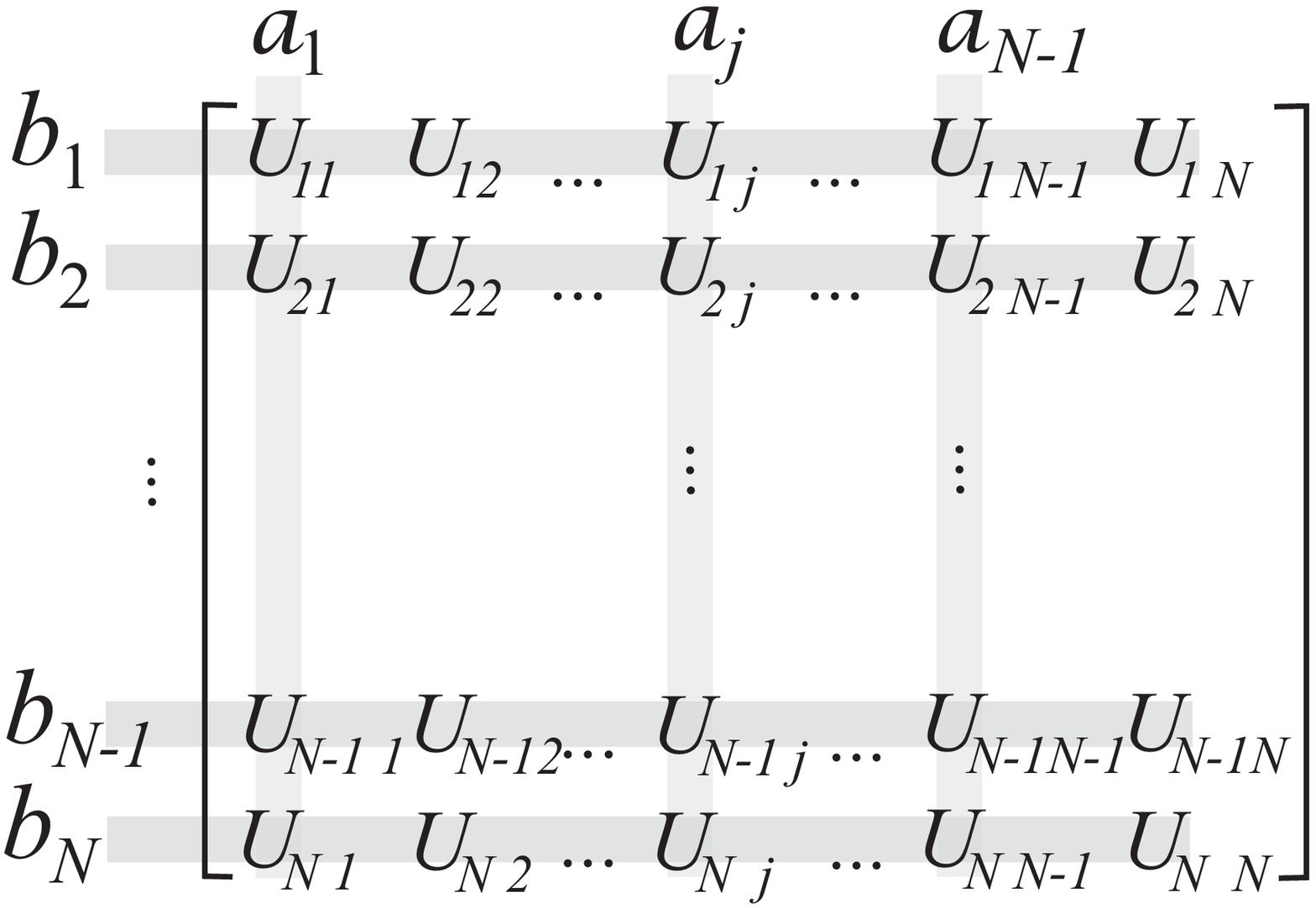}
    \caption{Relevant elements of a linear device transfer matrix
    when the number of input ports populated with signals are lesser than the total number of channels in transformation.
    The elements $U_{ij}$ that play a role in transformation~\eqref{eqn:linearTransform} in this specific case are those with $i=\overline{1,N}$ and $j=\overline{1,M}$.}
    \label{fig:fig1}
\end{figure}

Let us consider a special case of linear transformation, 
when the input is sparsely populated with signals,
i.e. 
the number of input ports excited by signals, $M$, is lesser than the total number of ports in the transformation: $M<N$.  
Therefore, not all elements of matrix $U$ are relevant for defining the transformation, so that the summation in \eqref{eqn:linearTransform} can  
be truncated to include only proper indices $j$,
which is illustrated in Fig.~\ref{fig:fig1}.
Taking the first $M$ input ports as signals,
the $N$-by-$M$ matrix constructed as a set of columns of the original $U_{ij}$
with  indices $i=\overline{1,N}$, $j=\overline{1,M}$. 
Henceforward, we use $\tilde{U}$ as a notation for this  $N$-by-$M$ matrix.
The parameter set for $\tilde{U}$ is now $2NM-M^2-N$, 
where extra $N$ has been subtracted as output phases is irrelevant for our consideration.

\begin{figure}
    \centering
    \includegraphics[width=0.45\textwidth]{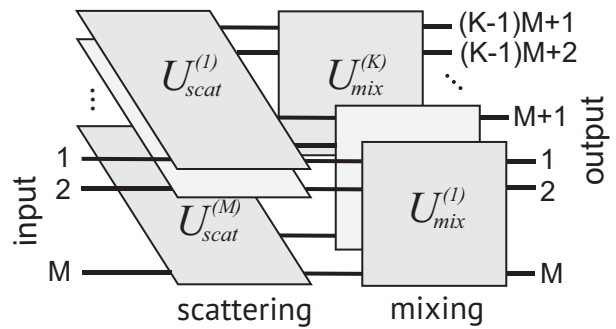}
    \caption{The architecture of an $N$-port interferometer
    operating on $M$ input signal ports
    for the specific case of $N=KM$, with $K$ being an integer. 
    Here, each of $M$ signals undergoes separate scattering among $K$ channels,
    which is followed by their mixing to give $N=KM$ output signals.
    The overall minimal depth of the interferometer is $K+M$.}
    \label{fig:fig2}
\end{figure}

Without loss of generality, we describe transformations with $N=KM$, where $K$ is a natural number. 
The underlying idea behind our method is to split the transform into two stages, as illustrated in Fig.~\ref{fig:fig2}, each having clear physical meaning.
In the first stage, 
each input signal undergoes separate scattering among $K$ ports,
during which it does not interfere with other signals.
Thus, there should be $M$ scattering transformations. 
Using the superscript to mark scattering transformation, 
and subscript to denote its output port,
the output field amplitude $b^{(m)}_i$  is related to the input $a_m$
by
    \begin{equation}\label{eqn:ScatteringTransform}
        b_i^{(m)}=c^{(m)}_ia_m,\quad(i=\overline{1,K}; m=\overline{1,M}),
    \end{equation}
where $c^{(m)}_i$ are complex-valued coefficients that define the scattering transform.
Since no losses are present, the coefficient of each scattering should obey $\sum_{i=1}^K{}|c^{(m)}_i|^2=1$.
In the matrix form, $m$-th scattering transforms is characterized by a unitary $K$-by-$K$ matrix $U^{(m)}_{scat}$  .
From~\eqref{eqn:ScatteringTransform} we readily have the first columns of these matrices known:  
    \begin{equation}\label{eqn:UscatCol}
        U_{scat}^{(m)}=\left(
        \begin{array}{cr}
            c^{(m)}_1\ldots\\
            c^{(m)}_2\ldots\\
            \vdots\\
            c^{(m)}_K\ldots
        \end{array}
        \right).
    \end{equation}
Obviously, the rest of the elements that are not written explicitly in \eqref{eqn:UscatCol} do not influence the scattering transforms and, therefore, they are can be chosen arbitrary,
provided that conditions imposed on unitary matrices are fulfilled.
The full explicit form of~\eqref{eqn:UscatCol} can be constructed, 
for example,  the iterative approach presented in~\cite{Jarlskog}, for example.
The overall scattering matrix acting upon  $M$ input signals, being written in a block-diagonal form, reads:
    \begin{equation}\label{eqn:ScatteringMatrix}
        U_{scat}=U_{scat}^{(1)}\oplus\ldots\oplus{}U_{scat}^{(M)},
    \end{equation}
and have $2KM-M-1$ parameters.
The vector of amplitudes at the input, $\mathbf{a}_{in}$,  
the combined transfer matrix~\eqref{eqn:ScatteringMatrix} acted on, is expressed
    \begin{equation}\label{eqn:inputamplitudes}
        a_{in,j}=a_m\delta_{j,1+(m-1)K},
    \end{equation}
$(m=\overline{1,M};j=\overline{1,N})$,
where $\delta_{mn}$ is the Dirac delta function.

In the second stage, the signal fields that have been distributed among multiple ports ($b_i^{(m)}$)
are brought to interfere with each other by mixing transformations,
that is characterized by unitary matrices $U_{mix}^{(k)}$ (see Fig.~\ref{fig:fig2}).
Obviously, to be potentially capable to perform as wide class of linear transformations as possible, every transformation $U_{mix}^{(k)}$ should be able to accept scattered fields from all $U_{scat}^{(m)}$, thus, their size is $M$-by-$M$.
In contrast to the scattering matrices that perform  far narrower class of transformation than the complete class of possible unitaries,
we assume that the mixing matrices can be configured to arbitrary unitary.
Therefore, to implement $U_{mix}^{(k)}$, one can use the previously proposed designs~\cite{Reck,Clements}. 
However, we emphasize the design by Clements~\cite{Clements}, because of the lower depth of the optical scheme, equal to the number of channels operated on,  
and robustness to losses.

We have the combined mixing transform matrix 
written in the block-diagonal form:
    \begin{equation}\label{eqn:MixingMatrix}
        U_{mix}=U_{mix}^{(1)}\oplus\ldots\oplus{}U_{mix}^{(K)},
    \end{equation}
having size $N$-by-$N$.
Notice that the order of channels in \eqref{eqn:ScatteringMatrix} and \eqref{eqn:MixingMatrix} are different (see Fig.~\ref{fig:fig2}).
In order to adjust the channel order of the scattering matrix~\eqref{eqn:ScatteringMatrix}
to the channel order of the  mixing matrix~\eqref{eqn:MixingMatrix},
we introduce the permutation matrix $P$
with elements: 
    \begin{equation}\label{eqn:Permut}
        P_{ij}=\delta_{1+m(i-1)K,j},\quad{}\text{where}\quad{}m:(m-1)K<i\le{}mK.
    \end{equation}

Therefore,  the transfer matrix of the overall device is the product of 
\eqref{eqn:ScatteringMatrix}, \eqref{eqn:Permut} and \eqref{eqn:MixingMatrix}:
    \begin{equation}\label{eqn:CombinedMatrix}
        U^{(sp)}=U_{mix}PU_{scat},
    \end{equation}
which has size $N$-by-$N$.
Next, to choose  elements of $U$ that are relevant for the case of input state under consideration use the occupation of the input ports~\eqref{eqn:inputamplitudes}.
Using \eqref{eqn:UscatCol}, \eqref{eqn:inputamplitudes}, \eqref{eqn:Permut} and \eqref{eqn:CombinedMatrix}, and
keeping only those columns of the explicitly-written matrix~\eqref{eqn:CombinedMatrix} that correspond to non-zero inputs, 
we finally derive the $N$-by-$M$ matrix:
    \begin{equation}\label{eqn:BlockMatrix}
        U^{(sp)}=\left(
        \begin{array}{c}
             B^{(1)}\\
             \hline
             B^{(2)}\\
             \hline
             \vdots\\
             \hline
             B^{(K)}
        \end{array}
        \right),
    \end{equation}
with 
    \begin{equation}
        \begin{split}
            &B^{(j)}=\\
            &\left(
            \begin{array}{cccc}
                c_j^{(1)}U^{(j)}_{mix,11} & c_j^{(2)}U^{(j)}_{mix,12} & \ldots & c_j^{(M)}U^{(j)}_{mix,1M}\\
                c_j^{(1)}U^{(j)}_{mix,21} & c_j^{(2)}U^{(j)}_{mix,22} & \ldots & c_j^{(M)}U^{(j)}_{mix,2M}\\
                \vdots & \vdots & \vdots & \vdots \\
                c_j^{(1)}U^{(j)}_{mix,M1} & c_j^{(2)}U^{(j)}_{mix,M2} & \ldots & c_j^{(M)}U^{(j)}_{mix,MM} 
            \end{array}
            \right)
        \end{split}
    \end{equation}
being an $M$-by-$M$  block comprising all elements of the mixing matrix $U_{mix}^{(j)}$  and  $j$-th coefficients $c_j^{(m)}$ for all $m$.
The field amplitudes at the output of the interferometer device 
are calculated with~\eqref{eqn:BlockMatrix}:
    \begin{equation}
        b_i=\sum_{j=1}^{M}U^{(sp)}_{ij}a_j.
    \end{equation}

Counting the number of quantities required to parametrize the 
scattering-mixing cascade~\eqref{eqn:Permut} gives $MN+K-M$,
which is lower than $2MN-M^2-N$ of the universal transformation,
suggesting non-universality of the modular architecture under study.
In particular, at large values of $M$ the parameter set of 
$U^{(sp)}$ is two times smaller than that of  $\tilde{U}$.

\section{Conclusion}

We have suggested a method for constructing a multichannel interferometer that implement a wide range of linear optical transforms, 
when the number of input ports occupied by signal fields are much lesser than the overall number of channels.
The modular architecture of  interferometers derived by this method is practically viable, since  each module can be created, tested and chosen separately.
Besides, if all the interferometer building blocks are allocated on a single monolithic substrate of a photonic chip, 
separation the scheme into connected modules can alleviate the hurdle of wiring with electronics.
Noteworthy,
the interferometers can be a combination of the free-space optics and 
integrated photonics chips, with the former implementing the scattering part  and 
the later performing the more technically demanding mixing part.

\section{Funding information}

Russian Science Foundation (RSF) No 17-72-10255.

\end{document}